\documentclass[aps,pre,reprint,superscriptaddress,showpacs,showkeys,twocolumn]{revtex4} 
\usepackage{amsfonts}
\usepackage{amssymb}
\usepackage{amsmath}
\usepackage{graphicx}
\usepackage{color}

\begin{document} 

\title{Quantifying magnetic anisotropy dispersion: Theoretical and experimental study of the magnetic properties of anisotropic FeCuNbSiB ferromagnetic films}

\author{T.~M.~L.~Alves} 
\affiliation{Instituto Federal de Educa\c{c}\~{a}o, Ci\^{e}ncia e Tecnologia do Rio Grande do Norte, 59015-000 Natal, RN , Brazil}
\affiliation{Departamento de F\'{i}sica Te\'{o}rica e Experimental, Universidade Federal do Rio Grande do Norte, 59078-900 Natal, RN, Brazil} 
\author{C.~G.~Bezerra} 
\affiliation{Departamento de F\'{i}sica Te\'{o}rica e Experimental, Universidade Federal do Rio Grande do Norte, 59078-900 Natal, RN, Brazil} 
\author{A.~D.~C.~Viegas} 
\affiliation{Insituto de F\'{i}sica, Universidade Federal do Rio Grande do Sul, 91501-970 Porto Alegre, RS, Brazil} 
\author{S.~Nicolodi} 
\affiliation{Insituto de F\'{i}sica, Universidade Federal do Rio Grande do Sul, 91501-970 Porto Alegre, RS, Brazil} 
\author{M.~A.~Corr\^{e}a} 
\affiliation{Departamento de F\'{i}sica Te\'{o}rica e Experimental, Universidade Federal do Rio Grande do Norte, 59078-900 Natal, RN, Brazil} 
\author{F.~Bohn} 
\email[Electronic address: ]{felipebohn@dfte.ufrn.br}
\affiliation{Departamento de F\'{i}sica Te\'{o}rica e Experimental, Universidade Federal do Rio Grande do Norte, 59078-900 Natal, RN, Brazil} 

\date{\today} 

\begin{abstract}
The Stoner-Wohlfarth model is a traditional and efficient tool to calculate magnetization curves and it can provides further insights on the fundamental physics associated to the magnetic properties and magnetization dynamics. Here, we perform a theoretical and experimental investigation of the quasi-static magnetic properties of anisotropic systems. We consider a theoretical approach which corresponds to a modified version of the Stoner-Wohlfarth model to describe anisotropic systems and a distribution function to express the magnetic anistropy dispersion. We propose a procedure to calculate the magnetic properties for the anisotropic case of the SW model from experimental results of the quadrature of magnetization curves, thus quantifying the magnetic anisotropy dispersion. To test the robustness of the approach, we apply the theoretical model to describe the quasi-static magnetic properties of amorphous FeCuNbSiB ferromagnetic films. We perform calculations and directly compare theoretical results with longitudinal and transverse magnetization curves measured for the films. Thus, our results provide experimental evidence to confirm the validity of the theoretical approach to describe the magnetic properties of anisotropic amorphous ferromagnetic films, revealed by the excellent agreement between numerical calculation and experimental results.
\end{abstract} 



\maketitle

\section{Introduction} 
\label{Introduction}

Magnetic properties and magnetization dynamics are fundamental issues for current and emerging technological applications and electronic devices. In the last decades, large efforts have been devoted to the investigation of soft magnetic materials. Among them, amorphous and nanocrystalline magnetic alloys have attracted great interest due to their soft magnetic properties~\cite{JMMM112p258, JAP64p6044}. In particular, amorphous FeCuNbSiB alloys, the precursor of the so-called Finemet nanocrystalline alloy, present remarkable soft magnetic features, evidenced by low coercive fields, high permeability, and high saturation magnetization~\cite{Davies_Gibbs}, making possible its employment in high performance and low energy consumption devices. Experimentally, several results have been obtained for FeCuNbSiB magnetic ribbons, and, consequently, the knowledgement about its magnetic properties and the crystallization process are well-established~\cite{JMMM112p258, JAP64p6044, IEEETM25p3324, JAP90p9186, APL83p2859}. However, although some reports found in literature have addressed distinct phenomena and aspects~\cite{JMMM272pE913,JMMM286p51,PB384p144,PB384p271,JAP101p033908,JAP104p033902,IEEETM44p3921,PSS8p070,JAP112p053910,JAP116p123903}, the complete understanding on the magnetic behavior of amorphous FeCuNbSiB ferromagnetic films is still lacking~\cite{Liu}.

From a theoretical point of view, modeling and numerical calculation of magnetization curves are the primary instruments for the investigation of magnetic structure, magnetization process, as well as other magnetic properties in magnetic systems~\cite{Stoner_Wohlfarth}. For instance, coercive field, remanent magnetization and magnetic anisotropy can be directly verified from the calculated curves. In this context, despite its simplicity, the Stoner-Wohlfarth (SW) model proposed in 1948~\cite{Stoner_Wohlfarth,IEEETM27p3475} is an efficient tool for calculating the magnetization curves and still represents an important approach to the understanding of the physical behavior of various types of ferromagnetic alloys~\cite{JAP53p2395}. In the traditional SW model, the magnetic properties of isotropic systems are investigated. However, it can be extended in order to take into account the presence of an effective uniaxial magnetic anisotropy and modelling anisotropic systems~\cite{JAP53p2395,IEEETM12p1015,JMMM345p147,JMMM328p53,Jap67p2881,JAP99p08Q504}, providing further insights on the fundamental physics associated to the magnetic properties and magnetization dynamics.

In this paper we report a theoretical and experimental investigation of the quasi-static magnetic properties of anisotropic systems. First of all, we consider a theoretical approach which corresponds to a modified version of the Stoner-Wohlfarth model to describe anisotropic systems and a distribution function to express the magnetic anistropy dispersion, and perform numerical calculations to verify the influence of the distribution function on the magnetization curves. We propose a procedure to calculate the magnetic properties for the anisotropic case of the SW model from experimental results of the quadrature of magnetization curves, thus quantifying the magnetic anisotropy dispersion. To test the robustness of the approach, we apply the theoretical model to describe the quasi-static magnetic properties of amorphous FeCuNbSiB ferromagnetic films. From experimental data, we perform calculations and directly compare theoretical results with longitudinal and transverse magnetization curves measured for the films. Thus, through the agreement between theory and experiment, we provide experimental evidence to confirm the validity of the theoretical approach proposed here to describe the magnetic properties of anisotropic amorphous ferromagnetic films.

\section{Experiment}
\label{Experiment}

Here, we investigate ferromagnetic films with nominal composition Fe$_{73.5}$Cu$_{1}$Nb$_{3}$Si$_{13.5}$B$_{9}$ and thicknesses of $100$ and $150$ nm. The films are deposited by magnetron sputtering onto glass substrates, covered with a $2$ nm-thick Ta buffer layer. The deposition process is performed with the following parameters: base vacuum of $10^{-7}$ Torr, deposition pressure of $5.2$ mTorr with a $99.99$\% pure Ar at $18$ sccm constant flow, and DC source with current of $65$ mA for the deposition of the Ta and FeCuNbSiB layers. During the deposition, the substrate with dimensions of $10 \times 4$ mm$^2$ is submitted to a constant magnetic field of $2$ kOe, applied along the main axis of the substrate in order to induce a magnetic anisotropy and define an easy magnetization axis. 

\begin{figure}[!h] 
\includegraphics[width=8.5cm]{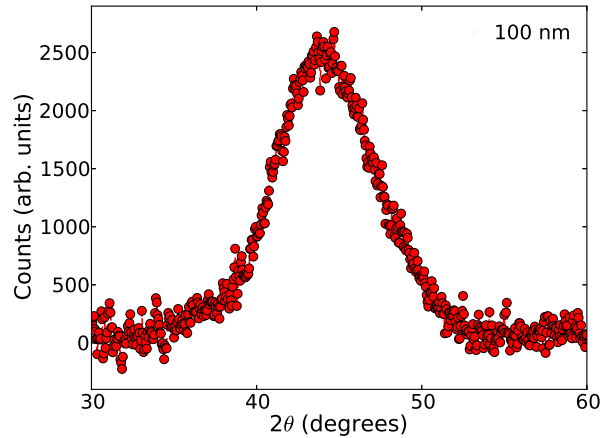}
\vspace{-.3cm}\caption{(Color online) High angle x-ray diffraction pattern for the FeCuNbSiB film with thickness of $100$ nm. The diffraction pattern confirms the amorphous character of the film. The thicker film presents similar behavior.} 
    \label{Fig_02} 
\end{figure}

While low-angle x-ray diffraction is employed to calibrate the film thickness, high-angle x-ray diffraction measurements are used to verify the structural character of the films. Figure~\ref{Fig_02} shows a high-angle x-ray diffraction for the FeCuNbSiB film with thickness of $100$ nm. Similar behavior is obtained for the thicker film. In particular, the pattern clearly indicates the amorphous state of the film, as depicted from the broad peak with low intensity, identified at $2\theta \sim 44.6^\circ$, and the absence of thin peaks with high intensity. Thus, the composition and the verified structural character infer that the studied films correspond to the amorphous precursor of the well-known Finemet~\cite{JAP64p6044, JMMM112p258}.

The magnetic properties are investigated through quasi-static magnetization curves. Longitudinal and transverse magnetization curves are obtained with a vibrating sample magnetometer and measured with the external magnetic field applied along and perpendicular to the main axis of the films, in order to verify the in-plane magnetic behavior. It is verified that when the field is applied in the plane of the film, the out-of-plane component of the magnetization is zero, indicating that the magnetization is in the plane, as expected due to the reduced thickness of the films. All the curves are obtained with maximum applied magnetic field of $\pm 300$ Oe, at room temperature.

\section{Theoretical approach} 
\label{Theoretical_approach}

Here, we focus on the investigation of the magnetic properties of ferromagnetic films, and perform numerical calculations of quasi-static magnetic curves. To this end, we employ a modified version of the Stoner-Wohlfarth model. Although in the original paper~\cite{Stoner_Wohlfarth,IEEETM27p3475} Stoner and Wohlfart have investigated isotropic systems, the SW model can be also employed to model anisotropic systems. The SW model is a very useful tool to obtain magnetization curves of heterogeneous alloys. It assumes that the system is composed by a set of monodomain size, non-interacting, and random oriented particles with uniaxial magnetic anisotropy. The reversal of the magnetization is due to coherent rotation of the magnetic particles, and thermal effects on magnetization are neglected~\cite{IEEETM27p3475}. 

\begin{figure}[!] 
    \begin{center} 
    \includegraphics[width=6.5cm]{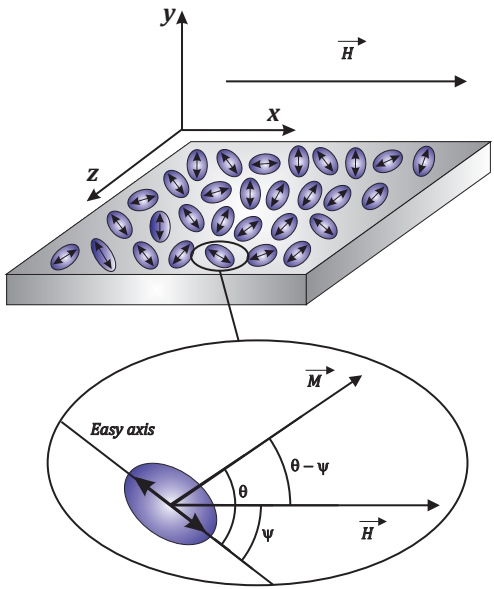} 
    \end{center} 
    \caption{Schematic diagram of the theoretical system of the SW model and the definitions of the magnetization and external magnetic field vectors considered for the numerical calculation of the magnetization curves. The magnetic system is composed by magnetic particles and the double arrows represent the direction of the magnetic anisotropy for each particle. We consider $\vec H$ as the external magnetic field vector, applied along the $x$ direction and in the system plane for all calculations, and $\vec M$ as the magnetization vector. Thus, in each particle, $\theta$ is the angle of $\vec M$ with respect to the effective magnetic anisotropy, while $\psi$ is the angle between $\vec H$ and the direction of the effective uniaxial magnetic anisotropy.} 
    \label{Fig_schematic_representation_01} 
\end{figure}

Figure~\ref{Fig_schematic_representation_01} presents the theoretical system described by the SW model and the definitions of the relevant vectors considered to perform the numerical calculations. Thus, from the appropriate magnetic free energy, a routine for minimization determines the values of the equilibrium angle $\theta_h$ of magnetization for a given external magnetic field, and we obtain the longitudinal and transverse magnetization curves.

When considering only the Zeeman interaction and the effective uniaxial anisotropy term in the SW model, the density of the magnetic free energy per particle can be written as~\cite{Stoner_Wohlfarth, Bertotti}
\begin{equation}
E(H,\theta,\psi)=-HM_S\cos(\theta-\psi)+K_{eff}\sin^2(\theta),
\label{Eq_energy_density}
\end{equation}
\noindent where $H$ is the external magnetic field, $\theta$ is the angle of the magnetization vector with respect to the effective magnetic anisotropy, $\psi$ is the angle between external magnetic field vector and the direction of the effective uniaxial magnetic anisotropy, $M_S$ is the saturation magnetization of the ferromagnetic material, which is considered to be uniform, and $K_{eff}$ is the effective uniaxial anisotropy constant. 

Considering reduced parameters, dividing Eq.~(\ref{Eq_energy_density}) by $2K_{eff}$, the reduced energy density $e(h,\theta,\psi)$ can be written as
\begin{equation}
e(h,\theta,\psi)=-h\cos(\theta-\psi)+{\frac {\sin^2(\theta)}{2}},
\label{Eq_reduced_energy_density}
\end{equation}
\noindent where $h = H / H_{Keff} $ is the reduced field, and $H_{Keff} = 2K_{eff} / M_S$ is the effective anisotropy field. 

The equilibrium state of the magnetization is represented by the equilibrium angle $\theta_h(\psi)$. It is obtained by minimizing Eq.~(\ref{Eq_reduced_energy_density}) for different values of $h$ and $\psi$~\cite{Stoner_Wohlfarth}, i.\ e., 
\begin{equation}
{\frac {\partial e(h,\theta,\psi)}{\partial \theta}}=0,
\end{equation}
\noindent whose solution $\theta_h(\psi)$ must satisfy the condition
\begin{equation}
{\frac {\partial^2 e(h,\theta,\psi)} {\partial^2 \theta} } > 0.
\end{equation}
\noindent Once $\theta_h(\psi)$ is determined, the magnetization $m(h)$ is usually calculated for two different directions, along and perpendicular to the magnetic field orientation. The respective components are defined as $m_{\parallel}(h)$ and $m_{\perp}(h)$~\cite{PB403p3563}, and can be written as
\begin{equation}
m_{\parallel}(h)=\cos[\theta_h(\psi)-\psi],
\label{Eq_longitudinal_magnetization}
\end{equation}
and
\begin{equation}
m_{\perp}(h)=\sin[\theta_h(\psi)-\psi].
\label{Eq_transverse_magnetization}
\end{equation}

The evolution of the magnetization $m$, as a function of the magnetic field $h$ for a given value of $\psi$, is governed by metastable states of the energy~\cite{Bertotti}. Since for some values of $h$ the energy density presents two possible stable states, the transition from one energy state to another, induced by the change of the magnetic field $h$, results in a critical and irreversible jump of the magnetization. The transition occurs at critical values $h_c(\psi)$ and $\theta_c(\psi)$, which can be found by solving the system of equations
\begin{equation}
\begin{array}{l}
\displaystyle {\frac {\partial e(h,\theta,\psi)}{\partial \theta}}=0, \\
\\
\displaystyle {\frac {\partial^2 e(h,\theta,\psi)} {\partial^2 \theta} } = 0.
\end{array}
\end{equation}
\noindent The $h_c(\psi)$ and $\theta_c(\psi)$ ​​values are considered to determine all equilibrium states $\theta_h(\psi)$ and, in the magnetization curve, provides the limit between the reversible and irreversilbe parts of the magnetization curve~\cite{JMMM345p147}.

\begin{figure}[!] 
\includegraphics[width=8.5cm]{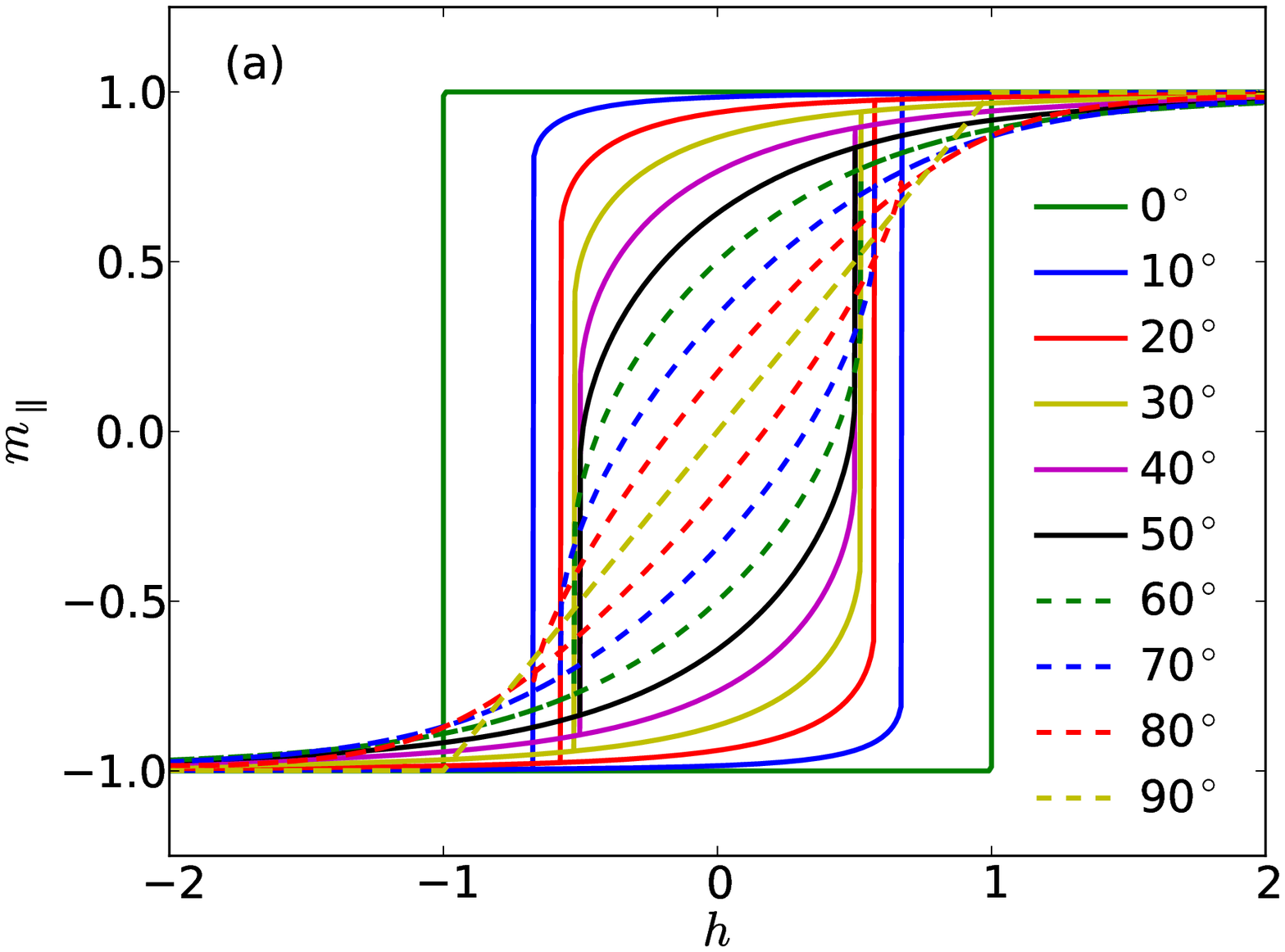}
\includegraphics[width=8.5cm]{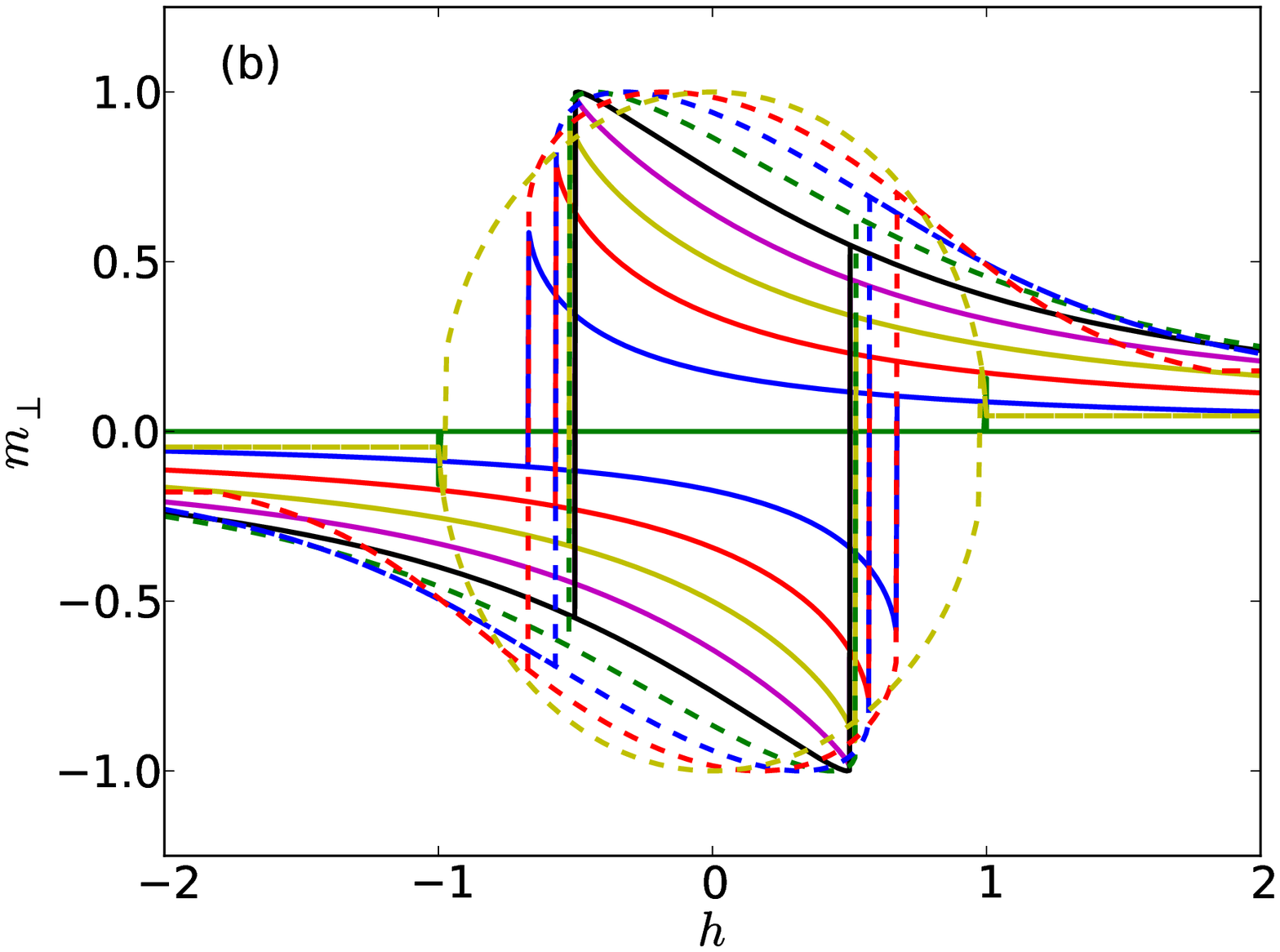}
\vspace{-.3cm}\caption{(Color online) Curves of the (a) longitudinal $m_{\parallel}$ and (b) transverse $m_{\perp}$ components of the magnetization as a function of the reduced magnetic field for different $\psi$ values, calculated for a single magnetic particle with an effective uniaxial magnetic anisotropy. } 
    \label{Fig_03} 
\end{figure}

To ilustrate the results calculated with the employed routine, Fig.~\ref{Fig_03} presents curves of the longitudinal and transverse components of the magnetization as a function of the reduced magnetic field for different $\psi$ values, calculated for a single magnetic particle with an effective uniaxial magnetic anisotropy. In this case, the $m_\parallel$ and $m_\perp$ curves present a clear dependence with the orientation between the easy magnetization axis and magnetic field, reflecting all traditional features of uniaxial systems~\cite{Cullity}, as expected.

For a set of magnetic particles whose components of magnetization per particle are given by Eqs.~(\ref{Eq_longitudinal_magnetization})-(\ref{Eq_transverse_magnetization}), the respective net values are obtained through the average values
\begin{equation}
\begin{array}{l}
\displaystyle <m_{\parallel}(h)>= {\frac {\int_{0}^{\pi/2} \cos(\theta_h(\psi)-\psi)f(\psi) \sin (\psi) d\psi} {\int_0^{\pi/2} f(\psi) \sin (\psi) d\psi} },
\end{array}
\label{Eq_magnetization_long}
\end{equation}
and
\begin{equation}
\begin{array}{l}
\displaystyle <m_{\perp}(h)>= {\frac {\int_{0}^{\pi/2} \sin(\theta_h(\psi)-\psi)f(\psi) \cos (\psi) d\psi} {\int_0^{\pi/2} f(\psi) \cos (\psi) d\psi}}.
\end{array}
\label{Eq_magnetization_transv}
\end{equation}
\noindent Here, $f(\psi)$ is a distribution function which describes the angular distribution, or the dispersion, of the magnetic anisotropy per particle. The limits of the integrals are $0^\circ$ and $90^\circ$ due to the symmetry of the uniaxial anisotropy.

As previously cited, the original SW model deals with the isotropic magnetic case, with no easy axis. A straight-forward generalization of the original SW model considers the presence of magnetic anisotropy. The more general anisotropic case has been also received attention and several very interesting works on this issue can be found in literature. This is the case of studies on the influence of the function $f(\psi)$ on the magnetic parameters of the magnetization curve~\cite{JMMM345p147,JMMM328p53}, description of systems with biaxial anisotropy~\cite{PRB85p134430,PhysRevB.88.094419}, with distribution of magnetic moment of particles~\cite{JMMM133p97} or of anisotropy strength~\cite{PRB68p104413,IEEETM37p2281,Jap67p2881,JAP99p08Q504}, as well as investigations of systems considering interactions between the particles through a mean field~\cite{PRB16p263,JMMM242p1093,JMMM323p2023,JMMM320pe73,Jap67p2881,JAP99p08Q504} or taking into account temperature effects~\cite{JMMM278p28,PRB72p054438,PRB75p184424,PRB88p094419}. On the other hand, just a few studies compare numerical calculations and experimental results, probably due to the fact that the determination of the appropriate magnetic free energy density can represent a hard task. In particular, some successful examples reside in the employment of the anisotropic SW model to study the influence of the magnetic anisotropy on the properties of sintered magnets composed by SmCo$_5$ and Co~\cite{JAP53p2395,IEEETM12p1015}, to describe transverse susceptibility of FeCoV thin films~\cite{JAP99p08Q504}, and to investigate the magnetoimpedance effect in ferromagnetic films~\cite{JAP110p093914,APL94p042501,JAP115p103908}.

In our approach, we consider the case of anisotropic SW model, $f(\psi)\neq 1$. The distribution function which describes the dispersion of magnetic anisotropy per particle may be a Gaussian $f(\psi)=\exp{(-\alpha{\psi}^2 )}$, a Lorentzian $f(\psi)={\beta}^2/({\beta}^2+{\psi}^2)$ or periodic functions, such as $f(\psi)={\cos^{n}(\psi)}$. 

In particular, we employ $f(\psi)={\cos^{n}(\psi)}$ since it has very similar behavior to that of the Gaussian function, with the advantage of being more convenient to perform analitycal~\cite{JMMM345p147} and numerical calculations. In this sense, the exponent $n$ defines how the distribution of the magnetic anisotropy is dispersed, i.~e., quantify the magnetic anisotropy dispersion. We also consider the parameter $\varphi$, ($0^\circ\le\varphi\le90^\circ$), which informs the location of the distribution. Thus, the distribution can be shifted from the origin, and the function can be written as
\begin{equation}
f(\psi)={\cos^{n}(\psi-\varphi)}.
\end{equation}

\begin{figure}[!] 
\includegraphics[width=8.5cm]{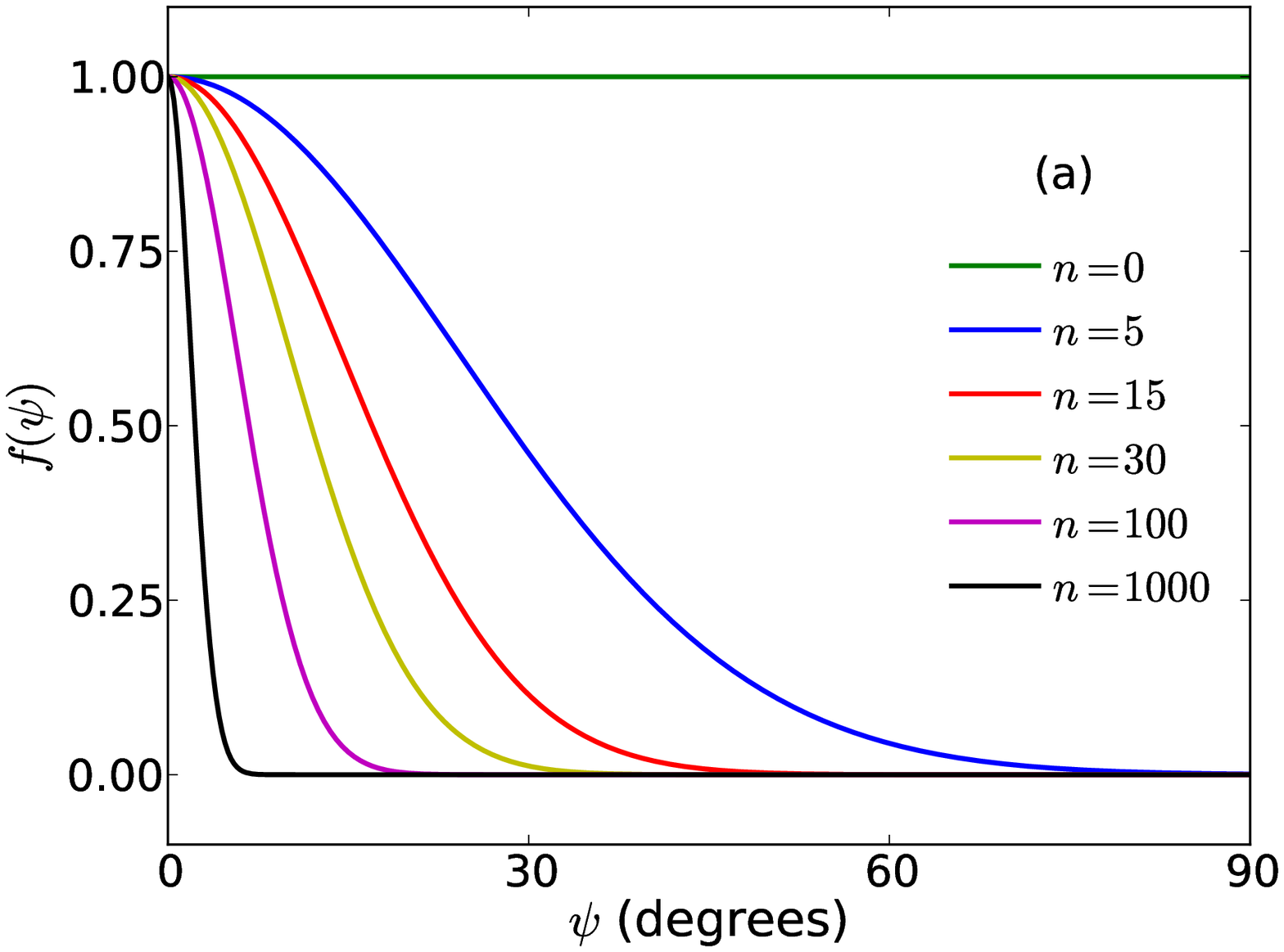}
\includegraphics[width=8.5cm]{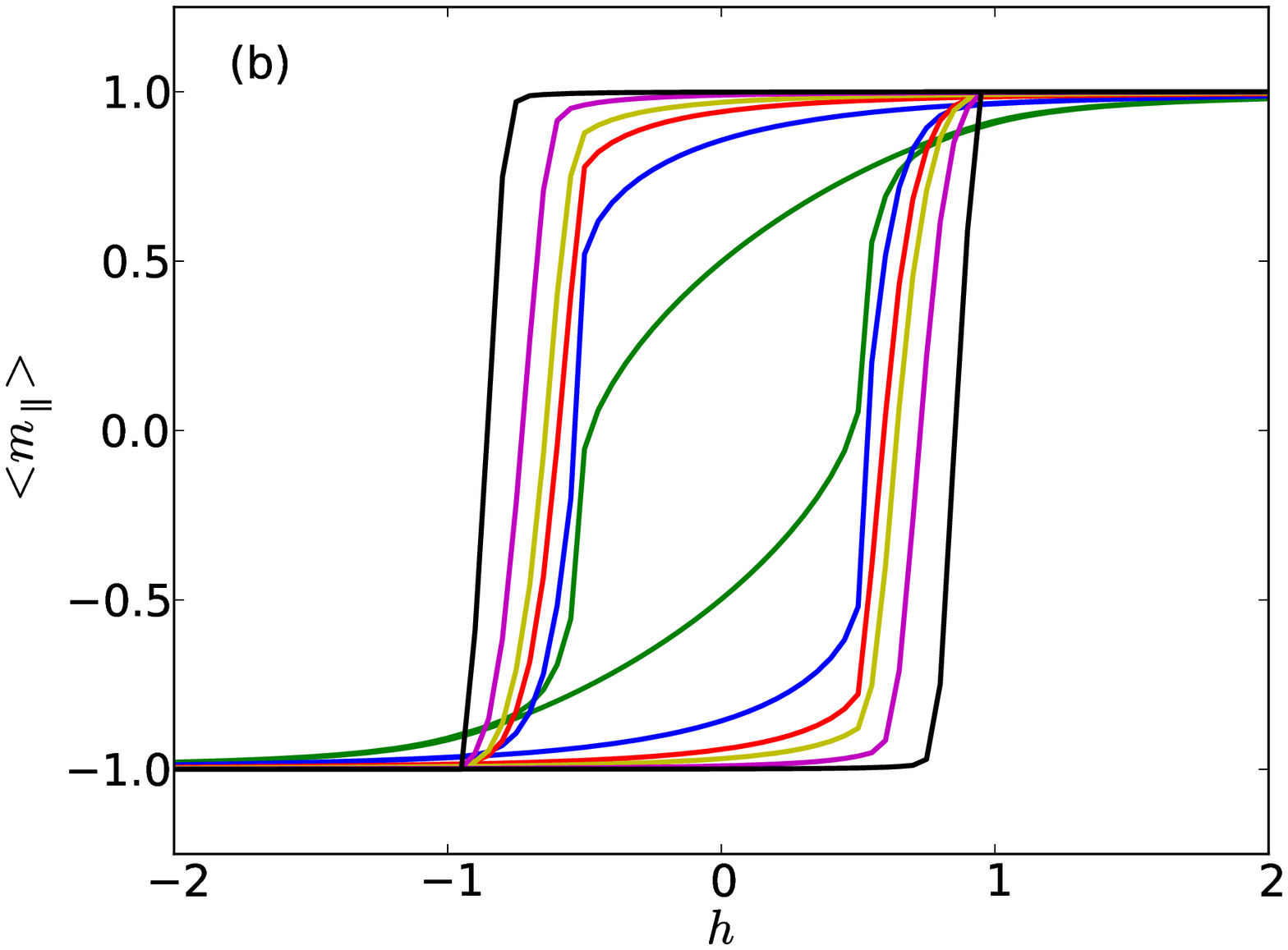}
\vspace{-.3cm}\caption{(Color online) (a) Distribution function $f(\psi)$ as a function of $\psi$ for different $n$ values, and (b) the corresponding curve of the net longitudinal $<m_{\parallel}>$ component of the magnetization per particle as a function of the reduced magnetic field. The numerical calculations are obtained considering $\varphi =0^\circ$. } 
    \label{Fig_04} 
\end{figure}

Figure~\ref{Fig_04} illustrates the influence of $n$ on the profile of the distribution function and on the corresponding curve of the net longitudinal component of the magnetization per particle as a function of the reduced magnetic field. The curves are calculated for different $n$ values with $\varphi=0^\circ$. For $n=0$, the isotropic SW model~\cite{Stoner_Wohlfarth} is recovered. As the $n$ value is increased, the dispersion of the anisotropy decreases and, consequently, the squareness, defined as the absolute value of $m_{\parallel}(h=0)$, of the magnetization curve increases. For larger $n$ values, the dispersion is very small and most of the particles has the magnetic anisotropy oriented along the same direction. It is worth to note that for $n=1000$, largest employed $n$ value corresponding to the presented case with the lowest anisotropy dispersion, the magnetization curve in Fig.~\ref{Fig_04}(b) is quite similar to the one with maximum squareness shown in Fig.~\ref{Fig_03}(a), obtained for a single magnetic particle with magnetic uniaxial anisotropy oriented along the external magnetic field. Moreover, it can be noted that the net magnetization, for magnetic field values close to the critical fields, does not rotate discontinuously as in the case of a single magnetic particle. This is a consequence of the dispersion of the magnetic anisotropy. The individual magnetic particles do rotate {\it discontinuously} for different $\psi$ values, particle by particle, as the magnetic field $h$ increases. However, the macroscopic net magnetization of the system {\it continuously} rotate towards the field direction. 

\begin{figure}[!] 
\includegraphics[width=8.5cm]{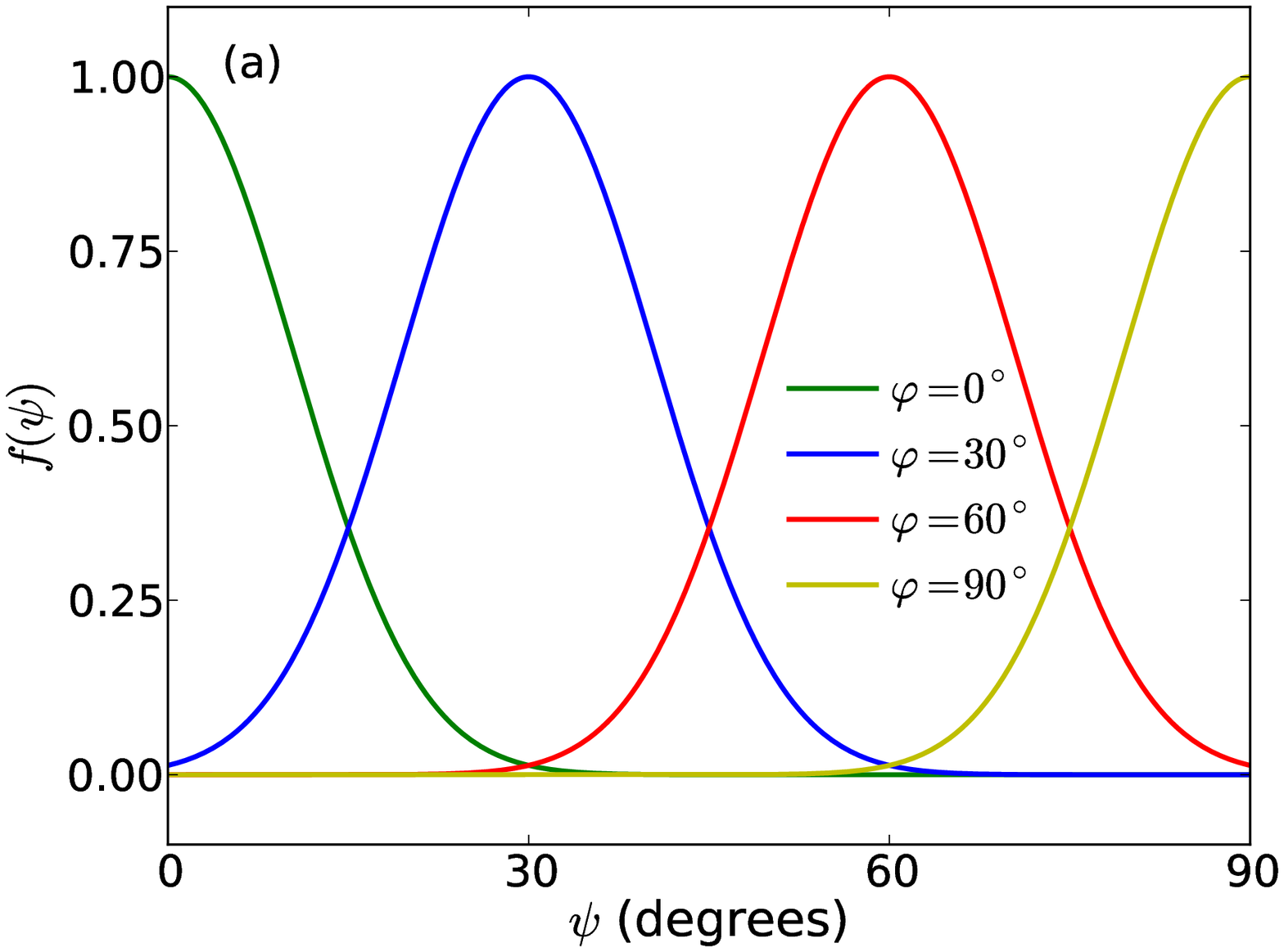}
\includegraphics[width=8.5cm]{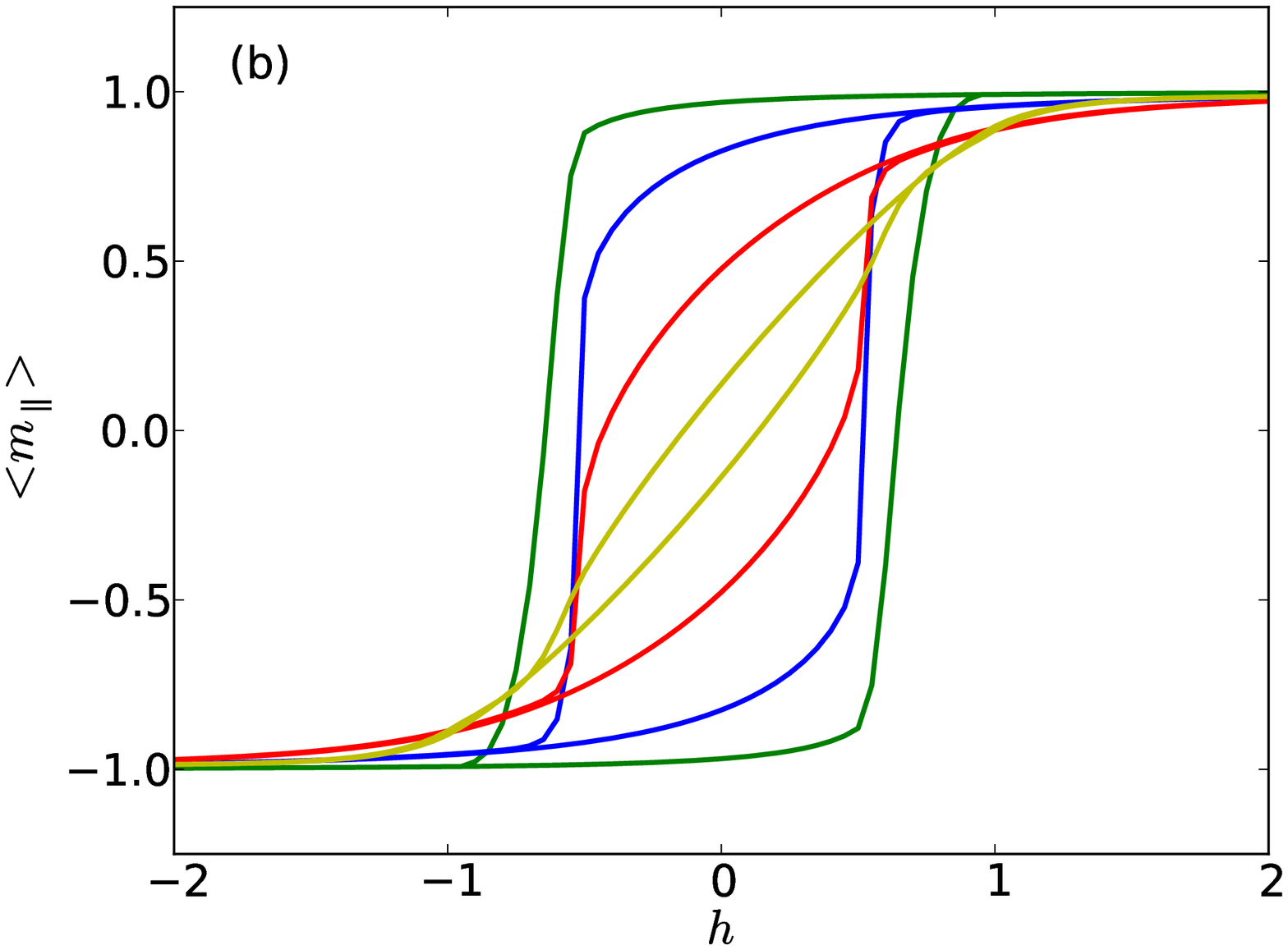}
\vspace{-.3cm}\caption{(Color online) (a) Distribution function $f(\psi)$ as a function of $\psi$ for different $\varphi$ values, and (b) the corresponding curve of the net longitudinal $<m_{\parallel}>$ component of the magnetization per particle as a function of the reduced magnetic field. The numerical calculations are obtained considering $n=30$.} 
    \label{Fig_05} 
\end{figure}

Figure~\ref{Fig_05} illustrates the influence of $\varphi$ on the profile of the distribution function and on the corresponding curve of the net longitudinal component of the magnetization per particle as a function of the reduced magnetic field. The curves are calculated for different $\varphi$ values with $n=30$. In this case, the general features observed in Fig.~\ref{Fig_03} are also reproduced here, considering a system with an effective uniaxial magnetic anisotropy with dispersion described by $f(\psi)$. In particular, the squareness of the magnetization curves and the coercive field are strongly changed as different $\varphi$ values are considered, a fact directly related to the change of the orientation between the center of the magnetic anisotropy dispersion and the external applied magnetic field.

\section{Comparison with the experiment} 
\label{Comparison_with_the_experiment}

The previous tests performed with our model have qualitatively described the main features of systems with effective uniaxial magnetic anisotropy. To confirm the validity of the theoretical approach, we investigate the quasi-static magnetic properties of amorphous FeCuNbSiB ferromagnetic films and compare the experimental results with numerical calculations. The complexity of the considered system, including different features previously studied, and the quantitative agreement with experimental results do confirm the robustness of our theoretical approach.

Although we have discussed the modifed version of the SW model in terms of magnetic {\it particles}, we point out that it has a wider applicability, since it does not require the existence of real particles. For instance, ferromagnetic films or certain portion of a continuous bulk system can also be studied by the same approach, as long as the assumption of uniform magnetization may be justified~\cite{Bertotti}.

For the studied samples, we understand each theoretical particle as a portion of the film with a given magnetization and uniaxial magnetic anisotropy. In this sense, the whole ferromagnetic film can be modeled as a set of magnetic particles, with the magnetization in the plane of the film, associated to a distribution function $f(\psi)$ describing the dispersion of effective magnetic anisotropy. Thus, we can directly perform the numerical calculation for the quasi-static magnetic properties from the model previously discussed.

\begin{figure}[!] 
\includegraphics[width=8.5cm]{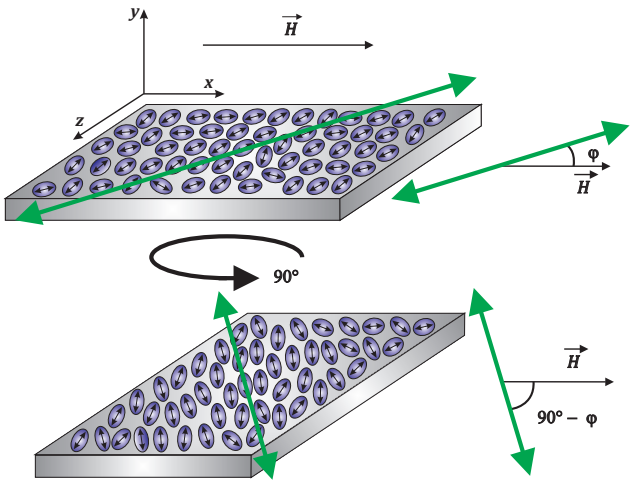}
\vspace{-.3cm}\caption{(Color online) Schematic diagram of the ferromagnetic film with effective magnetic anisotropy (Green arrow) with dispersion under the application of an external magnetic field. Numerical calculations of the net magnetization per particle obtained by applying the magnetic field along directions given by $\varphi$ and $(90^\circ -\varphi)$ from the center of the magnetic anisotropy dispersion.} 
    \label{Fig_06} 
\end{figure}

Experimentally, for each sample, the magnetization curves are measured with the external magnetic field applied along and perpendicular to the main axis of the film. In order to describe these experiments, we consider the calculated net magnetization per particle obtained by applying the magnetic field along directions respectively given by $\varphi$ and $(90^\circ -\varphi)$ from the center of the magnetic anisotropy dispersion, as illustrated in Fig.~\ref{Fig_06}. 
In this sense, $\varphi$ represents the fact that the exact angular orientation of the center of the magnetic anisotropy dispersion may be initially unknown. However, in some cases, such as when a constant magnetic field is used during the sample preparation, it is possible to roughly identify {\it a priori} the orientation of the magnetic anisotropy~\cite{IEEETM4p2281}. Thus, for these cases, the distributions can be, respectively, written as
\begin{equation}
f(\psi)_{\varphi}={\cos^{n}(\psi-\varphi)}
\label{Eq_dist_easy}
\end{equation}
and
\begin{equation}
f(\psi)_{(90^\circ -\varphi)}={\cos^{n}[\psi-(\pi/2 -\varphi)]}.
\label{Eq_dist_hard}
\end{equation}

The squareness of the longitudinal magnetization curves can be experimentally obtained from the observable $M_{r}/M_S$, where $M_r$ is the remanent magnetization. From Eq.~\ref{Eq_magnetization_long}, the squareness $<m_{\parallel}(h=0)>$ obtained to the same aforementioned experiments can be, respectively, expressed by
\begin{equation}
\displaystyle S_{\varphi} = {\frac {\int_0^{\pi/2} \cos(\theta_0(\psi)-\psi)f(\psi)_{\varphi} \sin (\psi) d\psi} {\int_0^{\pi/2} f(\psi)_{\varphi} \sin (\psi) d\psi} },
\label{Eq_squareness_0}
\end{equation}
and
\begin{equation}
\displaystyle S_{(90^\circ -\varphi)} = {\frac {\int_0^{\pi/2} \cos(\theta_0(\psi)-\psi)f(\psi)_{(90^\circ -\varphi)} \sin (\psi) d\psi} {\int_0^{\pi/2} f(\psi)_{(90^\circ -\varphi)} \sin (\psi) d\psi} }.
\label{Eq_squareness_90}
\end{equation}

Thus, from the squareness values, we can numerically determine the values of $n$ and $\varphi$ that best fit the experimental magnetization curves. The Eqs.~(\ref{Eq_squareness_0})-(\ref{Eq_squareness_90}) relate the parameters $n$ and $\varphi$, which describe the magnetic anisotropy dispersion, with the experimental squareness values obtained from the longitudinal magnetization curves measured with the external magnetic field applied along and perpendicular to the main axis of the films. For each magnetization curve, with a given squareness, we may determine a set of values ($n$, $\varphi$) that provides $<m_{\parallel}(h=0)>$ fitting the experimental data. In this case, the determination of a pair ($n$, $\varphi$), which univocally characterizes the ferromagnetic film, can be done by finding the intersection between the polynomial interpolation of the data obtained from the measurements with the field applied in both directions. 

\begin{figure}[!] 
\includegraphics[width=8.5cm]{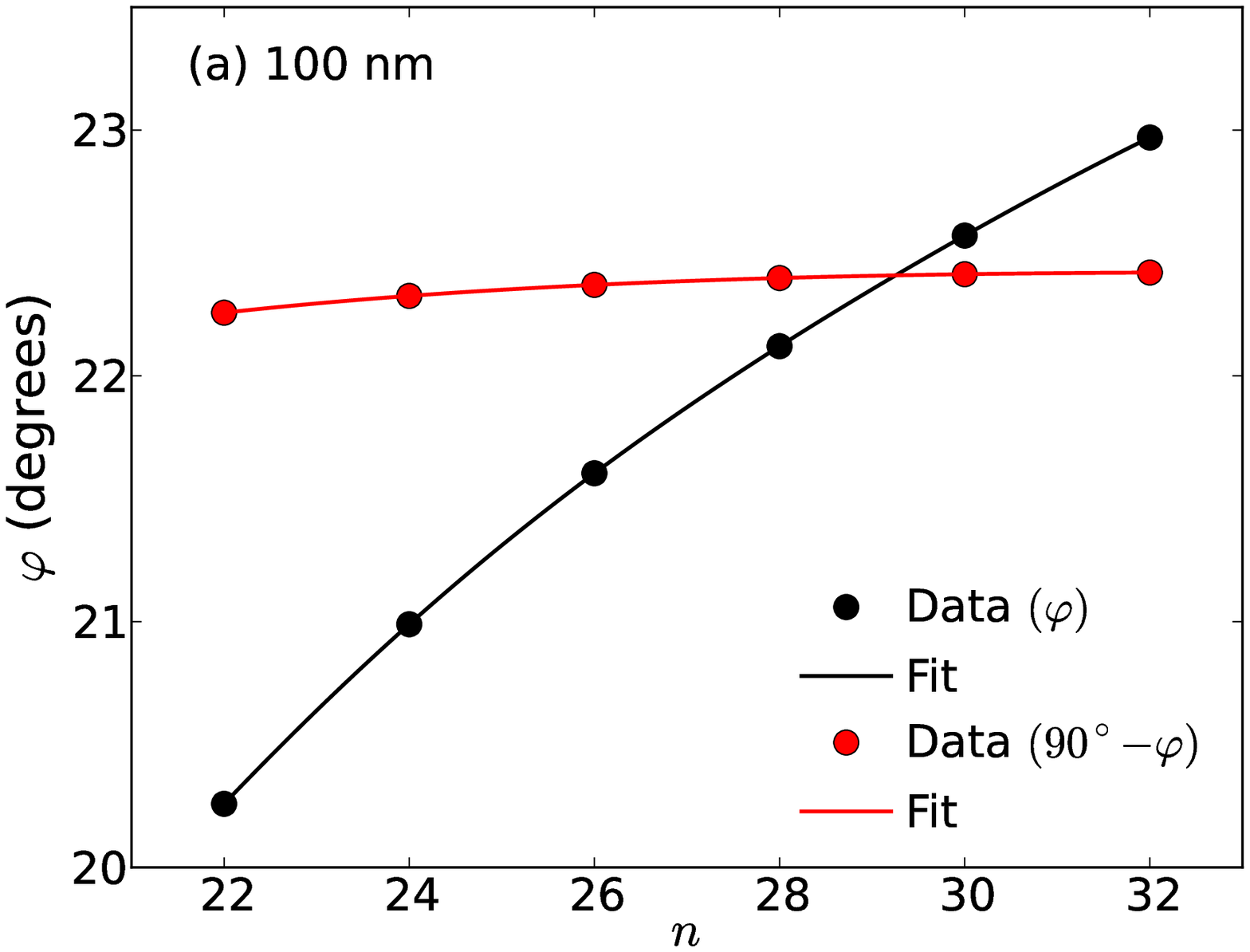}
\includegraphics[width=8.5cm]{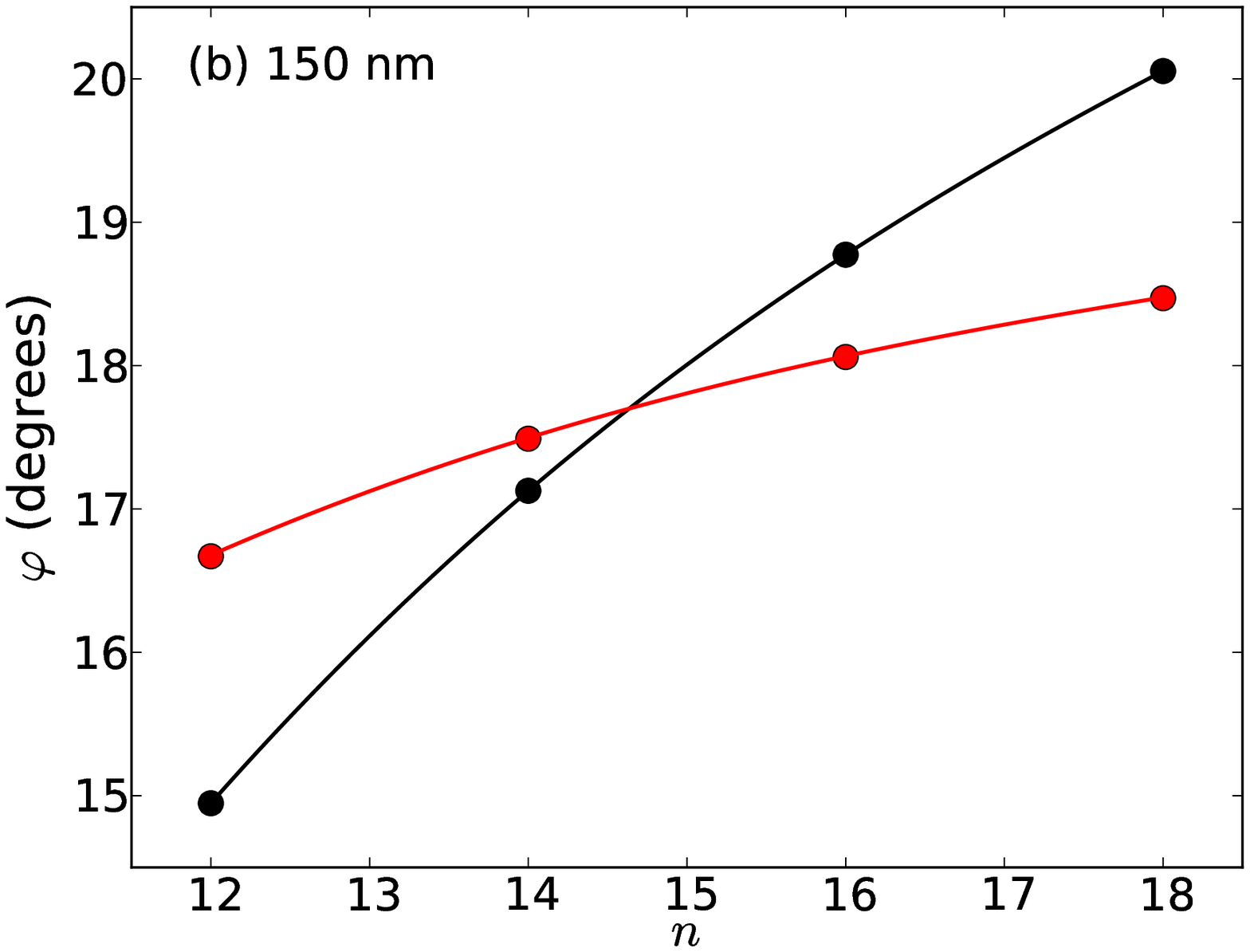}
\vspace{-.3cm}\caption{(Color online) Values of $\varphi$ as a function of $n$ that provides $<m_{\parallel}(h=0)>$ fitting the experimental data when the magnetic field is applied along a direction given by $\varphi$ and $(90^\circ -\varphi)$ from the center of the magnetic anisotropy dispersion for the ferromagnetic films with thicknesses of (a) $100$ and (b) $150$ nm. The solid lines are polynomial fittings of the discrete values. The intersection of the curves indicate the values of $n$ and $\varphi$ that provide the best fit of the experimental data.} 
    \label{Fig_07} 
\end{figure}

Figure~\ref{Fig_07} shows the sets of ($n$, $\varphi$) values obtained following the described procedure for the studied ferromagnetic films with different thicknesses. The summary of the physical parameters for both films are also reported in Table~\ref{tab1}.

\begin{table}
   \centering
   \setlength{\arrayrulewidth}{2\arrayrulewidth}
   \setlength{\belowcaptionskip}{0.5cm}
   \caption{Experimental values ​of squareness and the corresponding values of $n$ and $\varphi$ calculated for ferromagnetic films with distinct thicknesses.}
   \begin{tabular*}{\columnwidth}{@{\extracolsep{\fill}}cccccc}
      \hline\hline
Thickness & {$S_{\varphi}$} & $S_{(90^\circ -\varphi)}$ & $n$ & $\varphi \,(^{\circ} )$ & $H_K\,($Oe$)$ \\
(nm) & & & &  & \\
       \hline
 $100$ & $0.88$ & $0.37$  & $29.25$ & $22.37$ & $10.27$\\
 $150$ & $0.87$ & $0.33$  & $14.63$ & $17.66$ & $7.07$\\
      \hline\hline
   \end{tabular*}
   \label{tab1}
\end{table}

From the numerically obtained values of $\varphi$ and $n$, we are able to calculate the magnetization curves of the studied samples. In this case, the intensity of the magnetic anisotropy field $H_K$ is tuned in order to get the best fit of the experimental data. The obtained $H_K$ values are also shown in Table~\ref{tab1}. 

\begin{figure}[!] 
\includegraphics[width=8.5cm]{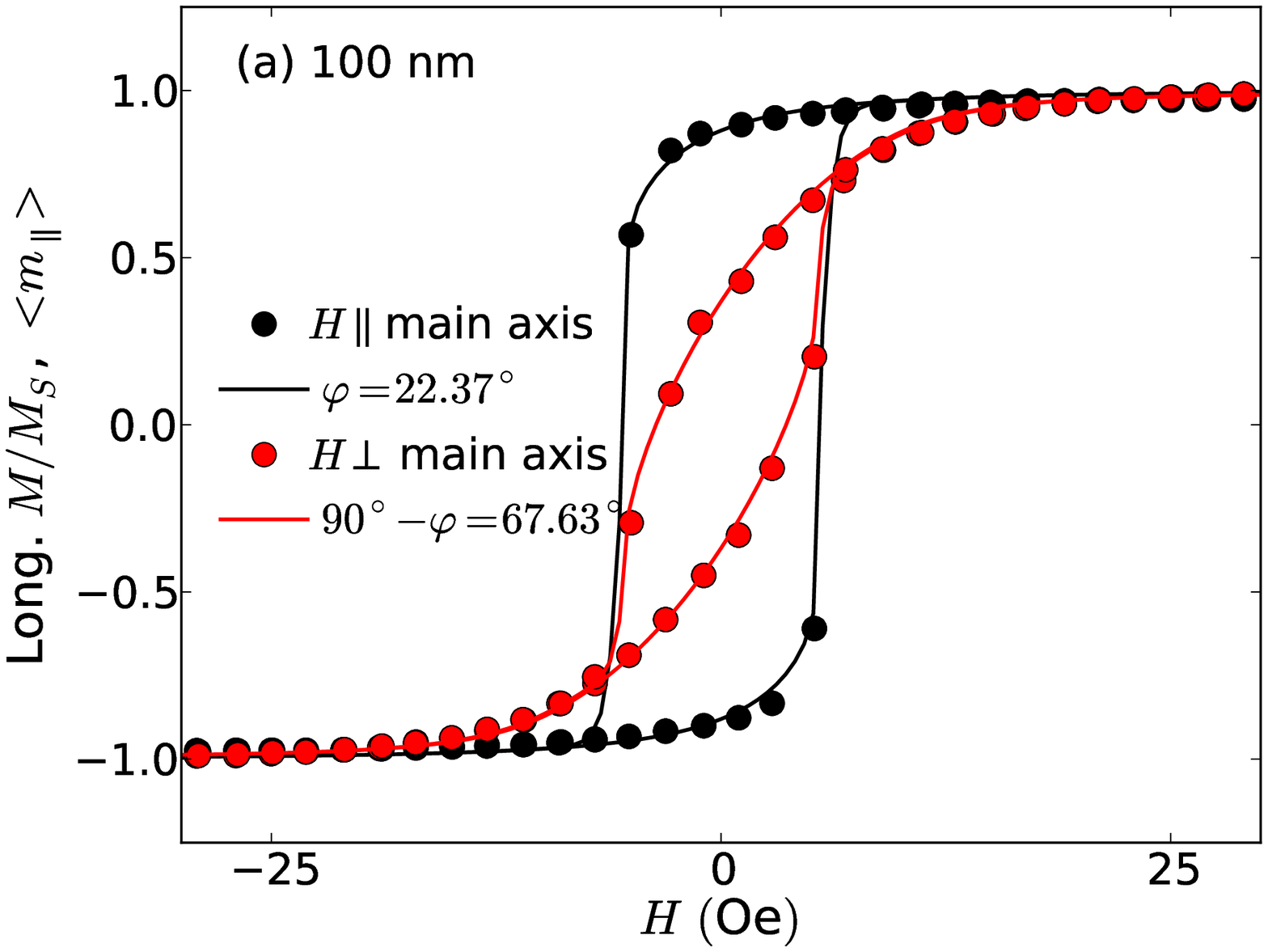}
\includegraphics[width=8.5cm]{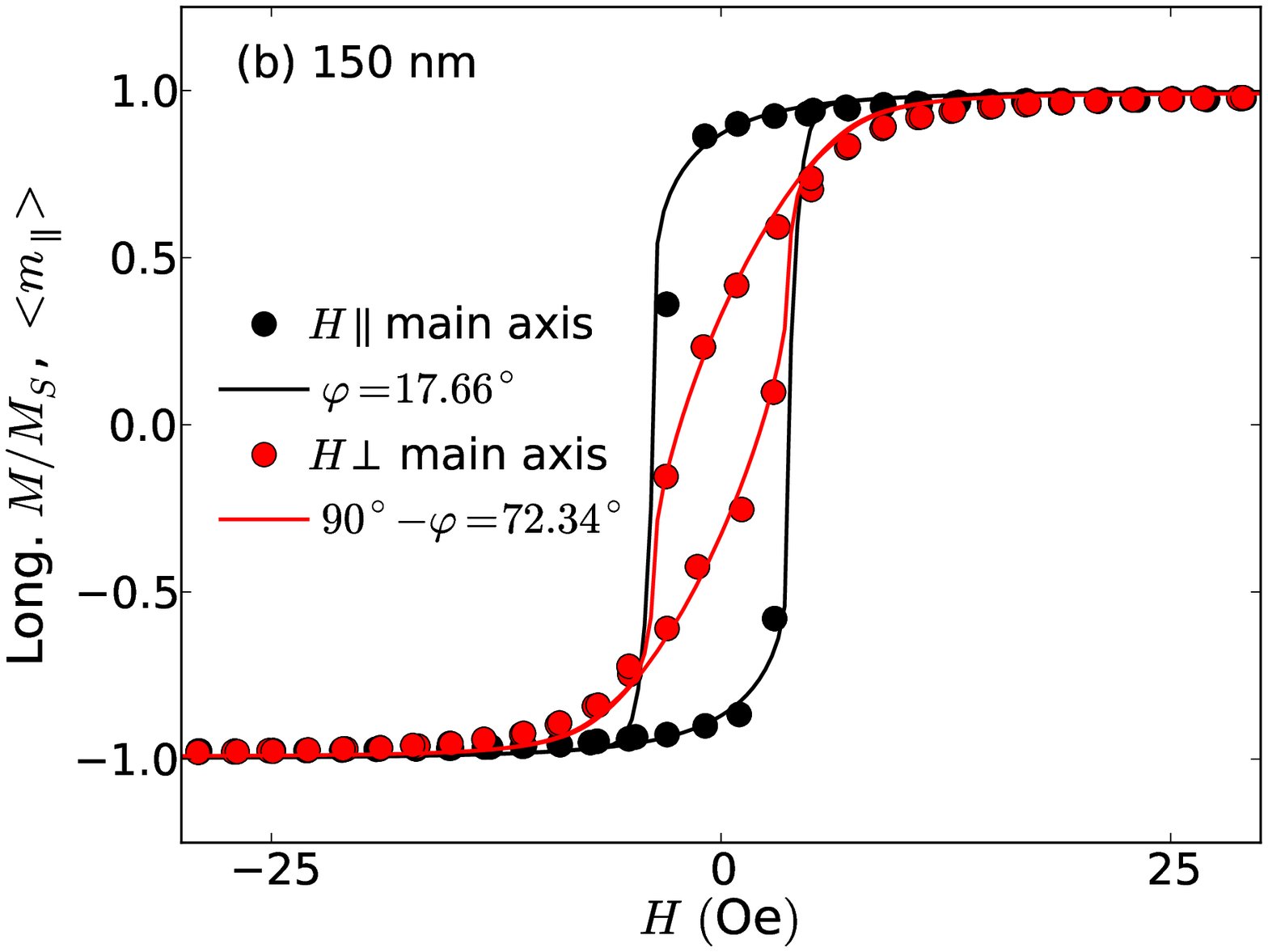}
\vspace{-.3cm}\caption{(Color online) Experimental and calculated results of the quasi-static magnetization curves for the films with thicknesses of (a) $100$ and (b) $150$ nm. Experimental longitudinal magnetization $M/M_S$ curves measured with the in-plane external magnetic field applied along (Black circles) and perpendicular (Red circles) to the main axis of the film, together with the numerical calculation of the net longitudinal $<m_\parallel>$ component of the magnetization per particle as a function of the external magnetic field $H$ obtained for $\varphi$ (Black line) and $(90^\circ - \varphi)$ (Red line). The numerical calculations are obtained considering the physical parameters summarized in Table~\ref{tab1}.} 
    \label{Fig_08} 
\end{figure}

Figure~\ref{Fig_08} shows experimental longitudinal magnetization curves measured with the in-plane external magnetic field applied along and perpendicular to the main axis of the film, together with the numerical calculation of the net longitudinal component of the magnetization per particle (Eq.~(\ref{Eq_magnetization_long})), as a function of the external magnetic field $H$, obtained for the two respective directions. Notice the striking quantitave agreement between experiment and theory. 

The angular dependence of the magnetization curves indicates an uniaxial in-plane magnetic anisotropy, induced by the magnetic field applied during the deposition process. As a matter of fact, by comparing experimental curves, it is possible to observe that the easy magnetization axis is oriented very close to the direction defined by the magnetic field applied during deposition. A small misalignment between them is a feature associated to stress stored in the film as the ferromagnetic layer thickness increases~\cite{JAP101p033908}, to the high positive magnetostriction of the alloy, to the rectangular shape of the substrate, as well as to a possible small deviation of the sample in the experiment during the measurement. Similar dependence of the magnetic behavior with the film thickness has been already observed and discussed in detail in Refs.~\cite{Magnetic_domains, PB384p144, JAP101p033908, JAP103p07E732, JAP104p033902}. Moreover, just considering the studied samples, we verify that the $H_K$, $n$ and $\varphi$ values decrease with the thickness. It indicates that the thicker film has higher magnetic anisotropy dispersion, a fact related to the local stress stored in the film as the thickness increases, as expected.

In order to confirm the robustness of the theoretical approach and obtain further information on the magnetic properties of the studied samples, we also consider tranverse magnetization curves. 

\begin{figure}[!ht] 
\includegraphics[width=8.5cm]{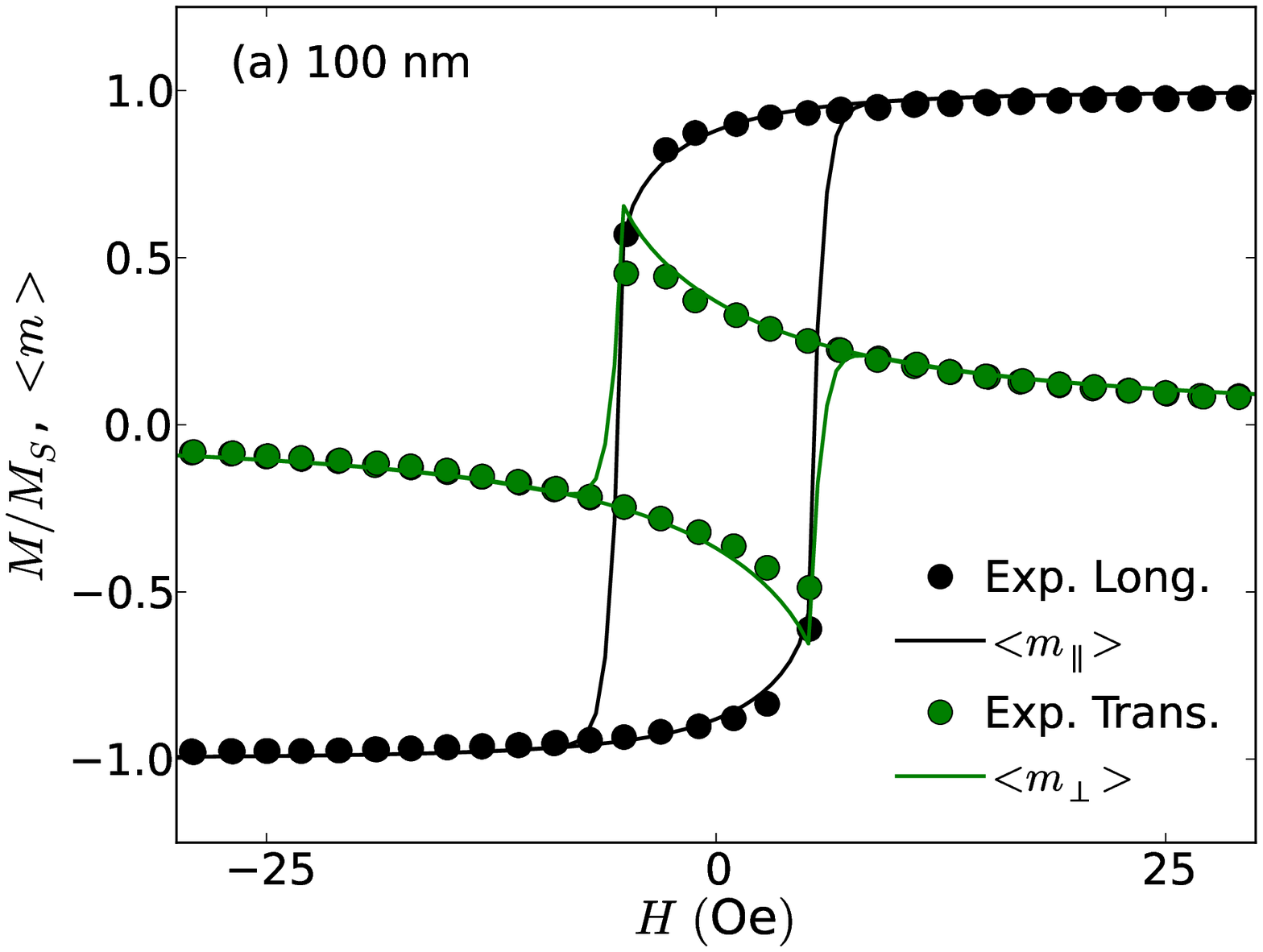}
\includegraphics[width=8.5cm]{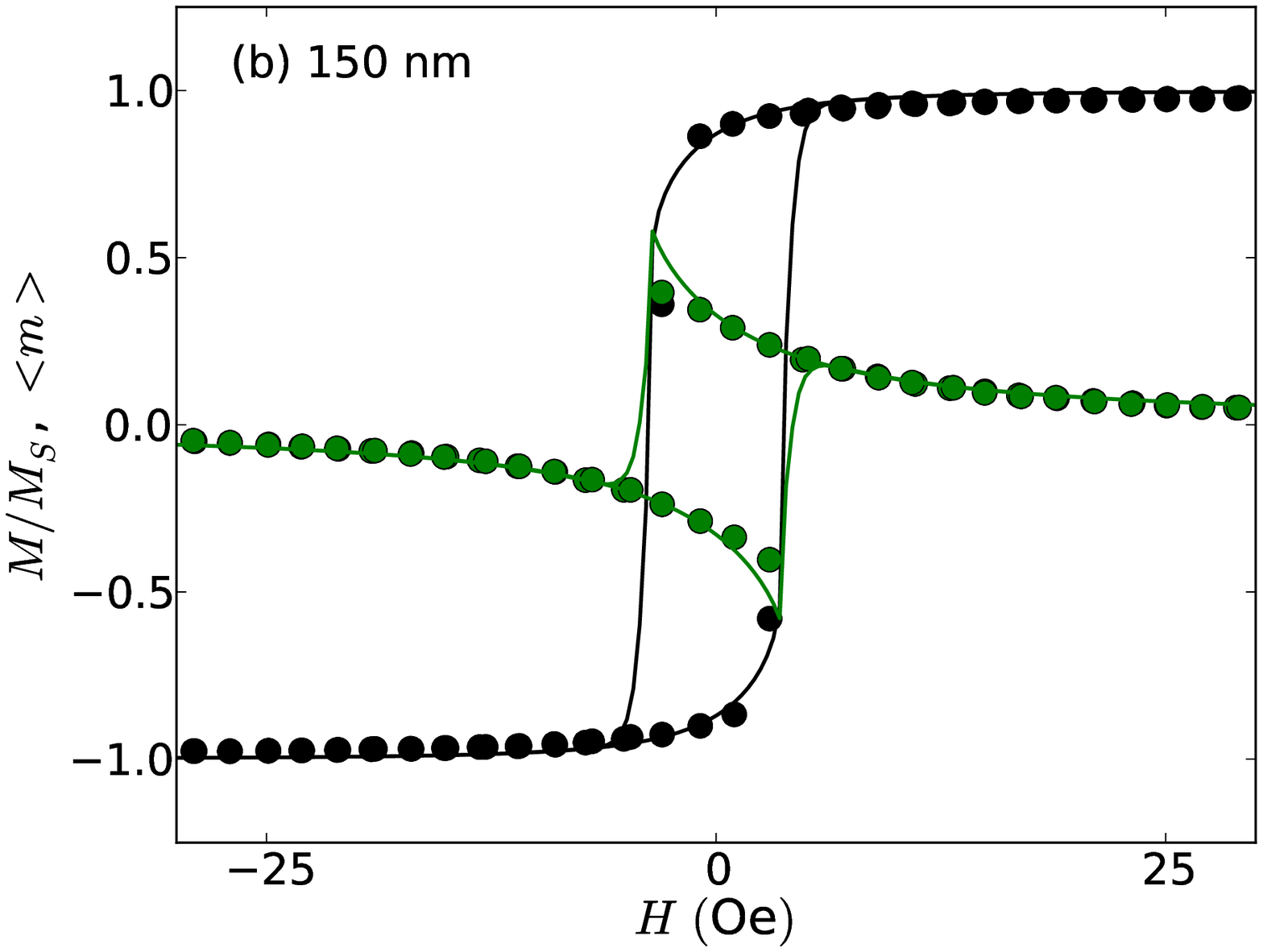}
\vspace{-.3cm}\caption{(Color online) Experimental and calculated results of the quasi-static magnetization curves for the films with thicknesses of (a) $100$ and (b) $150$ nm. Experimental longitudinal (Black circles) and transverse (Green circles) magnetization $M/M_S$ curves measured with the in-plane external magnetic field applied along the main axis of the film, together with the numerical calculation of the net longitudinal $<m_\parallel>$ (Black line) and $<m_\perp>$ (Green line) components of the magnetization per particle as a function of the external magnetic field $H$ obtained for $\varphi$. The numerical calculations are obtained considering the very same physical parameters summarized in Table~\ref{tab1}.} 
    \label{Fig_09} 
\end{figure}

\begin{figure}[!ht] 
\includegraphics[width=8.5cm]{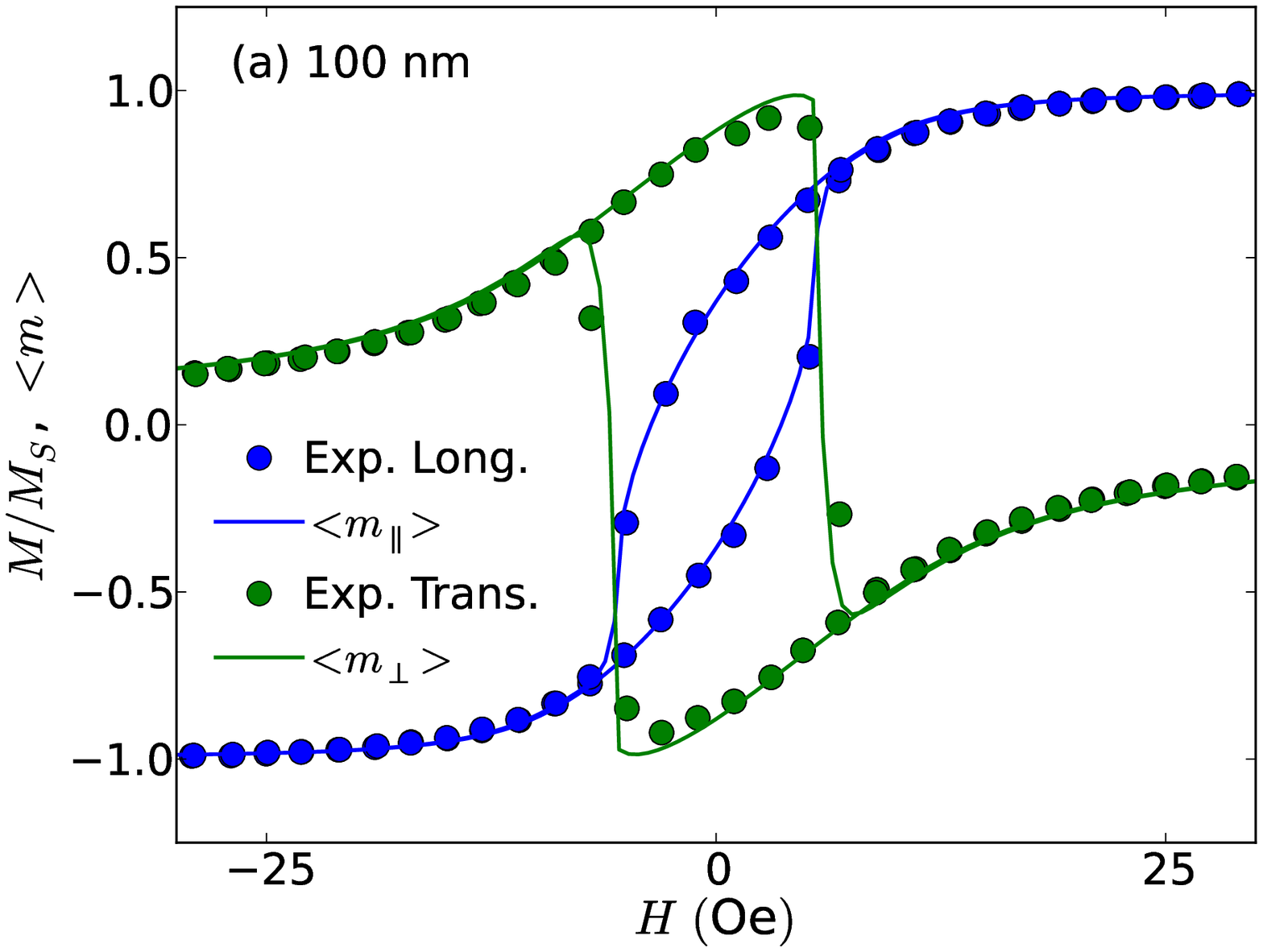}
\includegraphics[width=8.5cm]{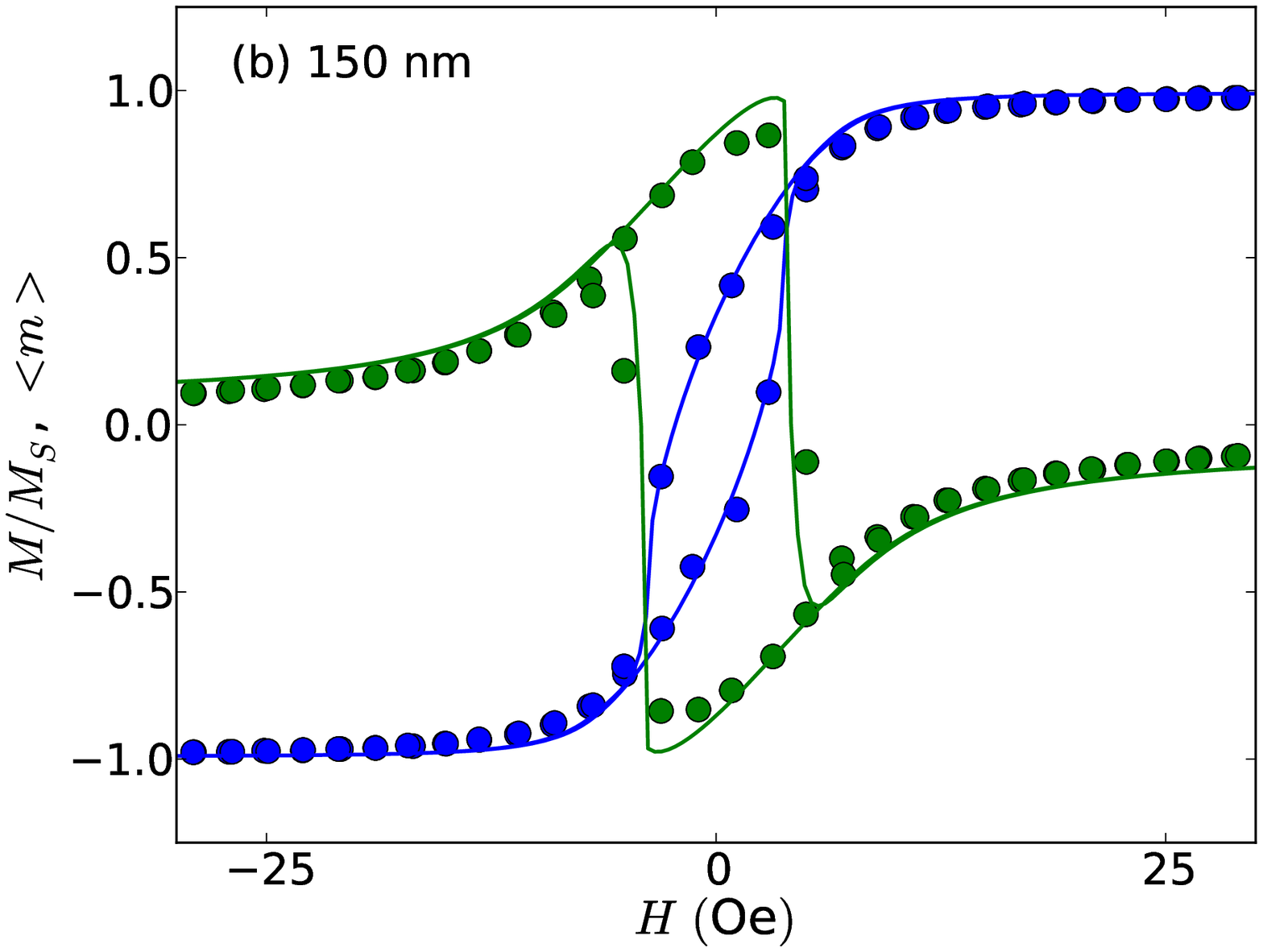}
\vspace{-.3cm}\caption{(Color online) Experimental and calculated results of the quasi-static magnetization curves for the films with thicknesses of (a) $100$ and (b) $150$ nm. Experimental longitudinal (Blue circles) and transverse (Green circles) magnetization $M/M_S$ curves measured with the in-plane external magnetic field applied perpendicular to the main axis of the film, together with the numerical calculation of the net longitudinal $<m_\parallel>$ (Blue line) and $<m_\perp>$ (Green line) components of the magnetization per particle as a function of the external magnetic field $H$ obtained for $(90^\circ - \varphi)$. The numerical calculations are obtained considering the very same physical parameters summarized in Table~\ref{tab1}.} 
    \label{Fig_10} 
\end{figure}

Figures~\ref{Fig_09} and \ref{Fig_10} show the experimental longitudinal and transverse magnetization curves measured with the in-plane external magnetic field applied along and perpendicular to the main axis of the film, respectively, together with the numerical calculation of the net longitudinal $<m_\parallel>$ and $<m_\perp>$ components of the magnetization per particle as a function of the external magnetic field $H$ obtained for $\varphi$ and $(90^\circ - \varphi)$. 

In particular, the numerical calculation is performed with the very same parameters fixed from the calculation of the aforementioned longitudinal magnetization curves. In this case, it is noticeable that they also present quantitave agreement with the experimental results, including shape and amplitude of all transverse magnetization curves. 

The numerical calculations performed with our theoretical approach present all classical features of the magnetization curves in systems with uniaxial anisotropy, including the magnetization behavior for distinct orientation between the anisotropy and external magnetic field, as well as the behavior with the magnitude of $H_K$, together with new features due to the anisotropy dispersion.

After all, we interpret the concordance between experiment and numerical calculations as a clear evidence that magnetic behavior of the amorphous FeCuNbSiB ferromagnetic films can be described by the modified version of the SW model employed here. Thus, we provide experimental support to confirm the validity of the theoretical approach.

\section{Conclusion}
\label{Conclusion}

In summary, in this paper we perform a theoretical and experimental investigation of the quasi-static magnetic properties of anisotropic systems. 

In particular, we consider a theoretical approach, which corresponds to a modified version of the Stoner-Wohlfarth model to describe anisotropic systems and a distribution function to express the magnetic anistropy dispersion. First of all, we perform numerical calculations to verify the influence of the distribution function on the magnetization curves. By considering the distribution function, we are able to insert naturally in the model some features observed in real samples, which are not verified considering a perfect uniaxial magnetic anisotropy, in concordance with experimental results found in literature. In this sense, although the determination of the appropriate magnetic free energy may represent a hard task, important physical parameters can be obtained.

At the same time, we propose a procedure to calculate the magnetic properties for the anisotropic case of the SW model from experimental results of the quadrature of magnetization curves. We perform experimental magnetization measurements in amorphous FeCuNbSiB ferromagnetic films. To interpret them, numerical calculations of the longitudinal and transverse magnetization curves are performed using the proposed theoretical approach. With the obtained physical parameters, we are able to describe the main features of the experimental magnetization curves. Thus, we provide experimental evidence to confirm the validity of the theoretical approach to describe the magnetic properties of ferromagnetic films, revealed by the excellent agreement between numerical calculation and experimental results. In this case, we infer that $n$ and $\varphi$ are important parameters that can be employed to quantify and describe the magnetic anisotropy dispersion.

Although we perform here all the analysis just for amorphous FeCuNbSiB films with distinct thicknesses, we understand that it can be extended and applied to any ferromagnetic system, for instance, any amorphous and nanocrystalline thin films. In this sense, the simplicity and robustness place this theoretical approach as a powerful tool to investigate the magnetic properties of soft ferromagnetic materials. Further investigations considering other ferromagnetic films with distinct magnetic properties are currently in progress.

\begin{acknowledgments} 
The research is partially supported by the Brazilian agencies CNPq (Grants No.~$471302$/$2013$-$9$, No.~$310761$/$2011$-$5$, No.~$476429$/$2010$-$2$, No.~$555620$/$2010$-$7$, No.~$307951$/$2009$-$0$, and No.~$482735$/$2009$-$0$), CAPES, FAPERJ, and FAPERN (Grant PPP No.~$013$/$2009$, and No.~$03/2012$). C.G.B, M.A.C.\ and F.B.\ acknowledge financial support of the INCT of Space Studies.
\end{acknowledgments}


\begin{thebibliography}{10}

\bibitem{JMMM112p258}
G.~Herzer,
\newblock J. Magn. Magn. Mater. {\bf 112}, 258  (1992).

\bibitem{JAP64p6044}
Y.~Yoshizawa, S.~Oguma, and K.~Yamauchi,
\newblock J. Appl. Phys. {\bf 64}, 6044 (1988).

\bibitem{Davies_Gibbs}
H.~A. Davies and M.~R.~J. Gibbs,
\newblock {\em Handbook of Magnetism and Advanced Magnetic Materials} (Wiley,
  2007).

\bibitem{IEEETM25p3324}
Y.~Yoshizawa and K.~Yamauchi,
\newblock IEEE Trans. Magn. {\bf 25}, 3324 (1989).

\bibitem{JAP90p9186}
M.~Ohnuma {\em et~al.},
\newblock J. Appl. Phys. {\bf 93}, 9186 (2003).

\bibitem{APL83p2859}
M.~Ohnuma, K.~Hono, T.~Yanai, H.~Fukunaga, and Y.~Yoshizawa,
\newblock Appl. Phys. Lett. {\bf 83}, 2859 (2003).

\bibitem{JMMM272pE913}
L.~Santi {\em et~al.},
\newblock J. Magn. Magn. Mater. {\bf 272–276}, E913  (2004).

\bibitem{JMMM286p51}
M.~Kerekes {\em et~al.},
\newblock J. Magn. Magn. Mater. {\bf 286}, 51  (2005).

\bibitem{PB384p144}
L.~Santi {\em et~al.},
\newblock Physica B {\bf 384}, 144 (2006).

\bibitem{PB384p271}
A.~Viegas {\em et~al.},
\newblock Physica B {\bf 384}, 271  (2006).

\bibitem{JAP101p033908}
A.~D.~C. Viegas {\em et~al.},
\newblock J. Appl. Phys. {\bf 101}, 033908 (2007).

\bibitem{JAP104p033902}
M.~Coisson {\em et~al.},
\newblock J. Appl. Phys. {\bf 104}, 033902 (2008).

\bibitem{IEEETM44p3921}
P.~Tiberto, F.~Celegato, M.~Coisson, and F.~Vinai,
\newblock IEEE Trans. Magn. {\bf 44}, 3921 (2008).

\bibitem{PSS8p070}
F.~Celegato, M.~Coisson, P.~Tiberto, F.~Vinai, and M.~Baricco,
\newblock Physica Status Solidi C {\bf 8}, 3070 (2011).

\bibitem{JAP112p053910}
M.~Coisson {\em et~al.},
\newblock J. Appl. Phys. {\bf 112}, 053910 (2012).

\bibitem{JAP116p123903}
F.~Zighem {\em et~al.},
\newblock J. Appl. Phys. {\bf 116}, 123903 (2014).

\bibitem{Liu}
Y.~Liu, D.~J. Sellmyer, and D.~Shindo,
\newblock {\em Handbook of Advanced Magnetic Materials} (Springer, 2006).

\bibitem{Stoner_Wohlfarth}
E.~C. Stoner and E.~P. Wohlfarth,
\newblock Philosophical Transactions of the Royal Society of London. Series A,
  Mathematical and Physical Sciences {\bf 240}, 599 (1948).

\bibitem{IEEETM27p3475}
E.~Stoner and E.~Wohlfarth,
\newblock IEEE Trans. Magn. {\bf 27}, 3475 (1991).

\bibitem{JAP53p2395}
C.~Searle, V.~Davis, and R.~Hutchens,
\newblock J. Appl. Phys. {\bf 53}, 2395 (1982).

\bibitem{IEEETM12p1015}
S.~R. Trout and C.~Graham,
\newblock IEEE Trans. Magn. {\bf 12}, 1015 (1976).

\bibitem{JMMM345p147}
M.~F. de~Campos, F.~A.~S. da~Silva, E.~A. Perigo, and J.~A. de~Castro,
\newblock J. Magn. Magn. Mater. {\bf 345}, 147  (2013).

\bibitem{JMMM328p53}
F.~A.~S. da~Silva, N.~A. Castro, and M.~F. de~Campos,
\newblock J. Magn. Magn. Mater. {\bf 328}, 53  (2013).

\bibitem{Jap67p2881}
R.~Iglesias, J.~M.~G. Merayo, and H.~Rubio,
\newblock Applied Physics Letters {\bf 67}, 2881 (1995).

\bibitem{JAP99p08Q504}
R.~Matarranz {\em et~al.},
\newblock Journal of Applied Physics {\bf 99}, 08Q504 (2006).

\bibitem{Bertotti}
G.~Bertotti,
\newblock {\em Hysteresis in Magnetism} (Academic Press, San Diego, 1998).

\bibitem{PB403p3563}
C.~Tannous and J.~Gieraltowski,
\newblock Physica B: Condensed Matter {\bf 403}, 3563  (2008).

\bibitem{Cullity}
B.~D. Cullity,
\newblock {\em Introduction to Magnetic Materials} (Addison-Wesley, New York,
  1972).

\bibitem{PRB85p134430}
A.~Tamion {\em et~al.},
\newblock Phys. Rev. B {\bf 85}, 134430 (2012).

\bibitem{PhysRevB.88.094419}
A.~Hillion {\em et~al.},
\newblock Phys. Rev. B {\bf 88}, 094419 (2013).

\bibitem{JMMM133p97}
A.~L. Ribeiro,
\newblock J. Magn. Magn. Mater. {\bf 133}, 97  (1994).

\bibitem{PRB68p104413}
S.~Wang, S.~S. Kang, J.~W. Harrell, X.~W. Wu, and R.~W. Chantrell,
\newblock Phys. Rev. B {\bf 68}, 104413 (2003).

\bibitem{IEEETM37p2281}
M.~Pasquale {\em et~al.},
\newblock IEEE Trans. Magn. {\bf 37}, 2281 (2001).

\bibitem{PRB16p263}
E.~Callen, Y.~J. Liu, and J.~R. Cullen,
\newblock Phys. Rev. B {\bf 16}, 263 (1977).

\bibitem{JMMM242p1093}
P.~Krivosik, C.~Appino, C.~Sasso, and M.~Pasquale,
\newblock J. Magn. Magn. Mater. {\bf 242-245}, 1093  (2002).

\bibitem{JMMM323p2023}
S.~Collocott,
\newblock J. Magn. Magn. Mater. {\bf 323}, 2023  (2011).

\bibitem{JMMM320pe73}
S.~Romero, M.~de~Campos, H.~Rechenberg, and F.~Missell,
\newblock J. Magn. Magn. Mater. {\bf 320}, e73  (2008).

\bibitem{JMMM278p28}
V.~Franco and A.~Conde,
\newblock J. Magn. Magn. Mater. {\bf 278}, 28  (2004).

\bibitem{PRB72p054438}
C.~de~Juli\'an~Fern\'andez,
\newblock Phys. Rev. B {\bf 72}, 054438 (2005).

\bibitem{PRB75p184424}
L.~He and C.~Chen,
\newblock Phys. Rev. B {\bf 75}, 184424 (2007).

\bibitem{PRB88p094419}
A.~Hillion {\em et~al.},
\newblock Phys. Rev. B {\bf 88}, 094419 (2013).

\bibitem{JAP110p093914}
M.~A. Corr\^{e}a {\em et~al.},
\newblock J. Appl. Phys. {\bf 110}, 093914 (2011).

\bibitem{APL94p042501}
R.~B. da~Silva {\em et~al.},
\newblock Appl. Phys. Lett. {\bf 94}, 042501 (2009).

\bibitem{JAP115p103908}
A.~M.~H. de~Andrade, M.~A. Corr\^{e}a, A.~D.~C. Viegas, F.~Bohn, and R.~L.
  Sommer,
\newblock J. Appl. Phys. {\bf 115}, 103908 (2014).

\bibitem{IEEETM4p2281}
M.~Pasquale {\em et~al.},
\newblock IEEE Trans. Magn. {\bf 37}, 2281 (2001).

\bibitem{Magnetic_domains}
A.~Hubert and R.~Sch{\"a}fer,
\newblock {\em Magnetic domains: The Analysis of Magnetic Microstructures}
  (Springer, New York, 1998).

\bibitem{JAP103p07E732}
N.~Amos {\em et~al.},
\newblock J. Appl. Phys. {\bf 103}, 07E732 (2008).

\end{thebibliography}
\end{document}